\documentclass[reprint,amsmath,nofootinbib,aip,jcp]{revtex4-1}

\usepackage{amssymb}
\usepackage{braket}
\usepackage{amsmath}
\usepackage{marvosym}
\usepackage{longtable}
\usepackage{braket}
\usepackage{xcolor,colortbl}
\usepackage{todonotes}
\definecolor{Gray}{gray}{0.85}
\definecolor{LightCyan}{rgb}{0.88,1,1}

\usepackage[left=1.25in,right=1.25in,top=1.0in,bottom=2.10in]{geometry}

\usepackage{blindtext}
\begin{document}

	\title{Full wave function optimization with quantum Monte Carlo
		and its effect on the dissociation energy of FeS}
	\author{Kaveh Haghighi Mood}
	\author{Arne L\"uchow}
	\email{luechow@pc.rwth-aachen.de}
	\affiliation{Institute of Physical Chemistry, RWTH Aachen University, Landoltweg 2, 52056 Aachen, Germany}	

\begin{abstract}
Diffusion quantum Monte Carlo calculations with partial and full optimization of the guide function are carried out for the
 dissociation of the FeS molecule. For the first time, quantum Monte Carlo orbital optimization for transition metal compounds is performed. It is demonstrated that energy optimization of the orbitals of a complete active 
 space wave function in the presence of a Jastrow correlation function is required to obtain agreement with the experimental 
 dissociation energy. Furthermore, it is shown that orbital optimization leads to a $^5\Delta$ ground state, in agreement
   with experiments, but in disagreement with other high-level ab initio wave function calculations which all predict 
   a $^5\Sigma^+$ ground state. The role of the Jastrow factor in DMC calculations with pseudo potentials is investigated.
   The results suggest that a large Jastrow factor may improve the DMC accuracy substantially at small additional cost.
\end{abstract}
\maketitle
\section{Introduction}
Understanding the chemistry of transition metal compounds is the key to the comprehension of phenomena like photosynthesis
 and the design of highly efficient catalysts. The iron sulfide molecule FeS is the smallest member of iron and sulfur 
 containing compounds which include important enzymes like nitrogenase. Despite its small size it is a very challenging 
 system for ab initio wave function methods.

Variational (VMC) and diffusion (DMC) quantum Monte Carlo methods~\cite{Wagner2014, Austin2012,Luechow2011} are the most
 dominantly used stochastic methods to study real materials in physics and chemistry when there is a necessity for accuracy
  beyond mean field theory. DMC can provide results at least as reliable as the quantum chemistry gold standard CCSD(T) 
  while being highly parallelizable on current hardware~\cite{grossman2002,Nemec2010,klahm20141,Klahm20147,Dubecky2016}. 

VMC and DMC methods have successfully been applied to transition metal compounds. 
Most calculations to date have employed a single determinant guide function with pseudopotentials, for instance 
the early calculation of low-lying states and the dissociation energy of TiC by one of the 
present authors~\cite{Sokolova2000}. 
Wagner and Mitas ~\cite{Wagner2007m,wagner2003} 
 investigated a series of 3d transition metal oxides using VMC and DMC while dissociation energies were computed 
 for transition metal carbonyl complexes by Diedrich et al.~\cite{Diedrich2005}. 
 Further work includes the 
  atomization and ionization energies for various vanadium oxide compounds~\cite{Bande2008}. 
  Doblhoff-Dier et al.~\cite{Doblhoff-Dier2016} studied the performance of DMC for the dissociation energy of a selected set of transition metal compounds.
  Antisymmetrized geminal product guide functions with a Jastrow factor were employed by
  Casula et al.~\cite{Casula2009} to describe the correct energy 
  ordering for the iron dimer with DMC whereas  
  Caffarel and coworkers studied the all-electron DMC spin density distribution for the $\text{CuCl}_2$
   molecule~\cite{Caffarel2014}. 

One of the present authors determined recently ionization and dissociation energies for 3d transition metal sulfides~\cite{Petz2011}.
 Despite the overall very good performance of DMC with single determinant guide functions, a large deviation from 
 experiment of 0.54 and 0.40 eV was observed for the dissociation energies of FeS and CrS, respectively. 
In this paper, FeS is therefore reexamined. We investigate in particular the effect of a multideterminant 
guide function and simultaneous Jastrow correlation function and orbital optimization on the quality of the DMC energies.      
 
\section{Methods}
Since there are comprehensive reviews on quantum Monte Carlo methods 
(QMC)\cite{HammondQMC,Luechow2011,Austin2012,Wagner2014}, VMC, DMC, and the form of the wave function 
employed in this work are discussed in this section only briefly.

\subsection{Variational quantum Monte Carlo}
In the variational quantum Monte Carlo method an upper bound to the ground state energy is calculated using a parametrized
 trial wave function $\Psi_t(\textbf{R},\textbf{p})$ with electron coordinates $\textbf{R}$ and parameters $\textbf{p}$
\begin{equation}
\begin{split}
E=\frac{\braket{\Psi_t(\textbf{R},\textbf{p})|H|\Psi_t(\textbf{R},\textbf{p})}}{\braket{\Psi_t(\textbf{R},\textbf{p})|\Psi_t(\textbf{R},\textbf{p})}}.\hspace{5cm}
\label{eq:smallu}
\end{split}
\end{equation}
$E$ can be estimated efficiently using Monte Carlo quadrature with importance sampling,
 \begin{equation}
  \begin{split}
  E =\int d\textbf{R} p(\textbf{R}) E_L(\textbf{R})=\lim\limits_{N\rightarrow \infty}\frac{1}{N}\sum_{i=1}^{N}E_L(\textbf{R}_i) \hspace{2cm}
  \end{split}
  \end{equation} 
where $E_L(\textbf{R})=H\Psi_t(\textbf{R})/\Psi_t(\textbf{R})$ is called local energy, and
 $p(\textbf{R})=\Psi^*_t(\textbf{R})\Psi_t(\textbf{R})/\braket{\Psi_t(\textbf{R})|\Psi_t(\textbf{R})}$ 
 is the probability distribution. The parameters $\textbf{p}$ are optimized with respect to the energy. 
 The quality of the VMC results depends strongly on the quality and flexibility of the trial wave function.
 
 \subsection{Diffusion quantum Monte Carlo}
 In DMC, the imaginary time Schr\"odinger equation with importance sampling transformation is employed~\cite{Reynolds82}
 \begin{equation}
 \begin{split}
- \frac{\partial}{\partial \tau} f(\textbf{R}.\tau)=-\frac{1}{2} \nabla^2f(\textbf{R},\tau)\\+\nabla\cdot(\textbf{b}(\textbf{R})f(\textbf{R},\tau))
+[E_L(\textbf{R})-E_t]f(\textbf{R},\tau)
\end{split}
 \end{equation}
  where $f(\textbf{R},\tau)=\Psi(\textbf{R}, \tau)\Psi_g(\textbf{R})$, $\Psi_g(\textbf{R})$ is the guide function, $\textbf{b}(\textbf{R})=\nabla \Psi_g/\Psi_g$ is called drift or quantum force, and $E_t$ is the trial or reference energy. The equation is solved stochastically in its integral form
  \begin{equation}
  \begin{aligned}
  f(\textbf{R}^\prime,\tau^\prime)=
  \int d\textbf{R}  K(\textbf{R}^\prime,\tau^\prime;\textbf{R},\tau)f(\textbf{R},\tau)
  \end{aligned}
  \end{equation}
 with the propagator $K$. In the short time limit $\Delta\tau = \tau' - \tau \to 0$ the propagator can be approximated by 
 the product of a drift diffusion propagator $K_d$ and a weighting/branching term $K_b$
     \begin{equation}
     K(\textbf{R}^\prime,\tau^\prime;\textbf{R},\tau)\approx K_d(\textbf{R}^\prime,\tau^\prime;\textbf{R},\tau)K_b(\textbf{R}^\prime,\tau^\prime;\textbf{R},\tau)  
     \end{equation}
    where
    \begin{equation}
    K_d(\textbf{R}^\prime,\tau^\prime;\textbf{R},\tau)=(2\pi\tau)^{-3N/2} e^{- \frac{(\textbf{R}^\prime-\textbf{R}-\Delta\tau \textbf{b}(\textbf{R}))^2}{2\tau}}  
    \end{equation} 
    and
    \begin{equation}
    K_b(\textbf{R}^\prime,\tau^\prime;\textbf{R},\tau)=e^{-\tau(E_L(\textbf{R}^\prime)+E_L(\textbf{R})-2E_t)/2} 
    \end{equation}
In practice, a sample is in turn propagated with $K_d$ and weighted/branched with $K_b$ while enforcing the nodes of $\Psi_g$ onto $\Psi$ (fixed node approximation). 
     The ground state energy is calculated with a mixed estimator~\cite{Reynolds82}. 
    After removal of the time step error by extrapolation $\Delta \tau \to 0$, the accuracy of the DMC energy depends only on the accuracy of the nodes of $\Psi_g$ which is a $3n-1$ dimensional hypersurface for $n$ electrons.

 \subsection{Wave functions and pseudo potentials}
 The same Slater Jastrow form is used for trial and guide function
   \begin{equation}
   \begin{split}
   \ket{\Psi}=e^U\ket{\Phi} 
   \end{split}
   \end{equation}
   where  $e^U$ is the symmetric Jastrow correlation factor describing the dynamic electron correlation, 
   and $\ket{\Phi}$ is an antisymmetric Slater determinant or a linear combination thereof. 
   $U(\boldsymbol{\beta})$ is expanded in many body terms depending on parameters $\boldsymbol{\beta}$. In this work, a 
   14-term Jastrow factor with electron-electron, electron-nucleus, and electron-electron-nucleus terms 
   in the Boys/Handy form as suggested for QMC by Schmidt and Moskowitz ~\cite{Schmidt1990} and a more accurate form with 
    cusp-less electron-electron-nucleus terms which is identified as sm666 (69 terms) in reference ~\cite{Luchow2015} are employed. 
   In the following, the two forms are denoted by J1 and J2, respectively. 
    
    The antisymmetric part $\ket{\Phi}$ is a 
    Slater determinant or a linear combination of configuration state functions (CSF) $\ket{\textbf{F}_i}$, spin adapted
     linear combination of Slater determinants,
    \begin{equation}
    \begin{split}
    \ket{\Phi}=\sum_{i=1}^{N_{csf}}c_i\ket{\textbf{F}_i} \hspace{3.8cm}
    \end{split}
    \label{eq:psi}
    \end{equation}
    
   \begin{equation}
   \begin{split}
   \ket{\textbf{F}_i}=\sum_{k}d_{i,k}\ket{\textbf{D}^\uparrow_k}\ket{\textbf{D}^\downarrow_k}. \hspace{2.8cm}
   \end{split}
   \label{eq:csf}
   \end{equation}
   Spin up and spin down determinants $\ket{\textbf{D}^\uparrow_k}$ and  $\ket{\textbf{D}^\downarrow_k}$ are built from orthogonal single particle molecular orbitals (MO) expanded in an atomic basis
         \begin{equation}
         \begin{split}
         \ket{\psi_k}=\sum_{m=1}^{N_{bas}}\alpha_{k,m}\ket{\chi_m}. \hspace{3.5cm}
         \end{split}
         \end{equation}
In this work, the triple zeta basis set by Burkatzki, Filippi, and Dolg has been employed
~\cite{Burkatzki2008,burkatzki2007-234105}. The corresponding small core relativistic pseudo potentials have been used to 
improve the efficiency of QMC and to include scalar relativistic effects. As usual in QMC, the nonlocal pseudo potentials 
are localized on a spherical grid using the trial wave function~\cite{Mitas1991}.

Jastrow parameters $\boldsymbol{\beta}$, CI coefficients $\textbf{c}$, and MO coefficients $\boldsymbol{\alpha}$ 
are variational parameters of the wave function. All parameters have been optimized with respect to the VMC energy. 
The MO coefficients are not optimized directly but rather using appropriate orbital rotations to keep the orthogonality
 of the orbitals~\cite{Toulouse2007}. To the best of our knowledge, this is the first time orbital optimization is 
 applied to transition metal compounds.

The perturbative and the linear method are used for wave function optimization, both of which make use of efficient statistical
 estimators for derivatives with respect to the parameters\cite{Toulouse2007}.
Special care has to be taken for the pseudo potential contribution to the local energy because the localized 
pseudo potential depends on all wave function parameters. Analytical derivatives with respect to all parameters
 have been implemented into the QMC code amolqc~\cite{Manten_JCP115_5362}, developed in our group.

Analytical derivatives of the localized pseudo potential are particularly expensive for the orbital rotations. Therefore,
 the perturbative method was employed for the MO optimization. In this method, only the energy denominator 
 $\Delta \epsilon_i$ requires derivatives of the pseudo potential (see Eq. (43) and (44) in Ref. \cite{Toulouse2007}). 
 Attempts to replace  $\Delta \epsilon_i$ by guess values were not successful. Instead, we calculated $\Delta \epsilon_i$ 
 with low accuracy using a small subset of the sample and fixed the energy denominators for subsequent iterations.
       
Due to the complete active space (CAS) nature of the Slater part of the wave function, many spin-up and spin-down
 determinants in the CSF expansion are identical and calculated only once. The 630 Slater determinants of the CAS wave 
 function for the FeS $^5\Sigma^+$ are built from only nine different spin-up and 126 spin-down determinants. This leads to
  substantial speed up compared to a naive calculation of all Slater determinants~\cite{Scemama2016}.
  The multideterminant DMC calculations in this work require only about four times the CPU time for corresponding 
  single determinant calculation.

\section{Results}
FeS has been studied with many experimental and theoretical techniques. The ground state was assigned $X\ ^5\Delta$ by 
Zhang et el. \cite{Zhang1996} using mass spectrometry. This assignment was firmly established by Takano et al. 
using microwave spectroscopy~\cite{Takano2004}. The dissociation energy is not very accurately known. Drowart et 
al.~\cite{DrowartJ67} determined $D_0 = 3.31(15)$ eV following an earlier measurement of an upper bound of 3.34 eV by 
Marquart and Berkowitz~\cite{Marquart1963}. An experimental bond length of 2.017 {\AA} is obtained for the $X\ ^5\Delta$ 
state from the rotational constant measured by Takano and coworkers~\cite{Takano2004}. DeVore et al. 
performed IR spectroscopy on FeS in different matrices and obtained a harmonic frequency of 540 cm$^{-1}$ while 
Zhai et al.~ reported 520(30) cm$^{-1}$ employing photo electron spectroscopy\cite{DeVore1975,Zhai2003}.  
Using dispersed fluorescence spectroscopy, Wang et al.~ obtained more accurately 518 $\pm$ 5 cm$^{-1}$\cite{Wang2011}.

This molecule has also been studied extensively using density functional theory (DFT) and wave function methods. 
Surprisingly for such a small molecule, there is disagreement about the ground state.
Many DFT calculations correctly predict a $^5\Delta$ ground state~\cite{Hubner1998,Bridgeman2000,Schultz2005a,Wu2007,Clima2007,Liang2009,Li2013}, 
but some functionals including B3LYP and VSXC predict a $^5\Sigma^+$ ground state~\cite{Hubner1998, Wu2007, Schultz2005a} 
with several basis sets.
High-level multireference calculations resulted all in a $^5 \Sigma^+$ ground state. Sauer and coworkers used CASSCF/ACPF 
and found the $^5\Sigma^+$ state to lie 0.14 eV below the $^5 \Delta$ state~\cite{Hubner1998}. Clima and Hendrickx employed
 CASPT2 and also found a $^5\Sigma^+$ ground state~\cite{Clima2007}. An earlier CASSCF/ICACPF calculation by Bauschlicher 
 and Maitre considered only the $^5\Delta$ state~\cite{Bau1995}.

In this work, DMC with single determinant and CAS multideterminant guide functions is used to determine the ground state
symmetry, the dissociation energy, and the spectroscopic constants.
The electronic configuration for the $^5\Delta$ state is $\sigma^2\pi^4\sigma^2\delta^3\sigma^1\pi^2$ and
 $\sigma^2\pi^4\sigma^2\delta^2\sigma^2\pi^2$ for $^5\Sigma^+$. 
In the CASSCF calculations, 12 electrons were distributed in the active space formed by the 3d and 4s orbitals of the 
iron atom and the 3p orbitals of the sulfur atom. 
Dissociation energies are calculated by subtracting the atomic energies in their ground states $^3P$ and $^5D$ for sulfur
 and iron, respectively. State averaged CASSCF calculations were carried out for the atoms to preserve symmetry.
Initial wave functions for QMC were created using Hartree-Fock (HF), B3LYP, and CASSCF with the GAMESS code~\cite{Gamess}. 
The QMC code amolqc\cite{Manten_JCP115_5362} developed in our group is used for all QMC calculations.
              
\subsection{Effect of orbital optimization in single determinant QMC calculations}

VMC and DMC energies for single determinant wave functions were calculated for 
HF, B3LYP Kohn-Sham orbitals, and energy-optimized orbitals
for both the $^5\Delta$ and the $^5\Sigma^+$ state. The bond lengths were fixed at the minima obtained from the potential 
curves calculated at the multideterminant DMC level (see below), 2.031 {\AA} for $^5\Delta$ and 2.000 {\AA} for
 $^5\Sigma^+$. The Jastrow parameters were optimized with respect to the energy in all calculations. If both Jastrow 
 parameters and orbitals were optimized the parameters were optimized simultaneously. The DMC results, extrapolated to zero 
 time step, are shown in Table~\ref{TAB:SingleConf}.

\begin{table}
    \caption{FeS single determinant DMC energies in $E_h$ with HF, B3LYP, and VMC-optimized (opt orbs) orbitals and both Jastrow
        factors (Jas). All Jastrow parameters are optimized. Parentheses indicate 
        one standard deviation.    
    }
     \begin{tabular}{lllll}
        \hline\hline
        state        & Jas     & HF             & B3LYP             & opt orbs\\
        \hline
        $^5\Delta$   & J1   & -134.0181(4)   & -134.0426(4)      & -134.0426(4)\\ 
        $^5\Sigma^+$ & J1   & -134.0163(4)   & -134.0406(4)      & -134.0386(4)\\
        \hline
        $^5\Delta$   & J2  &  -134.0234(4)  & -134.0486(4)      & -134.0489(4)\\ 
        $^5\Sigma^+$ & J2  &  -134.0223(4)  & -134.0460(4)      & -134.0456(3)\\
                \hline\hline    
    \end{tabular}
    \label{TAB:SingleConf}
\end{table}

\begin{table}
    \caption{FeS single determinant VMC energies in $E_h$ with HF, B3LYP, and VMC-optimized (opt orbs) orbitals and both Jastrow
        factors (Jas). All Jastrow parameters are optimized.    
    }
    \begin{tabular}{lllll}
        \hline\hline
        state     & Jas   & HF             & B3LYP             & opt orbs\\
        \hline
        $^5\Delta$   & J1  &  -133.8975(4)  &  -133.9164(4)     & -133.9270(4)\\ 
        $^5\Sigma^+$ & J1  &  -133.8961(4)  &  -133.9107(4)     & -133.9225(4)\\
        \hline
        $^5\Delta$   & J2 & -133.9451(4)   &  -133.9708(4)   &  -133.9754(3) \\ 
        $^5\Sigma^+$ & J2 & -133.9440(4)   &  -133.9664(4)   &  -133.9681(3) \\   
        \hline\hline     
    \end{tabular}
    \label{TAB:SingleConfVMC}
\end{table}
  
\begin{figure}
     \centering
     \begin{minipage}{.48\linewidth}
         \includegraphics[width=\linewidth]{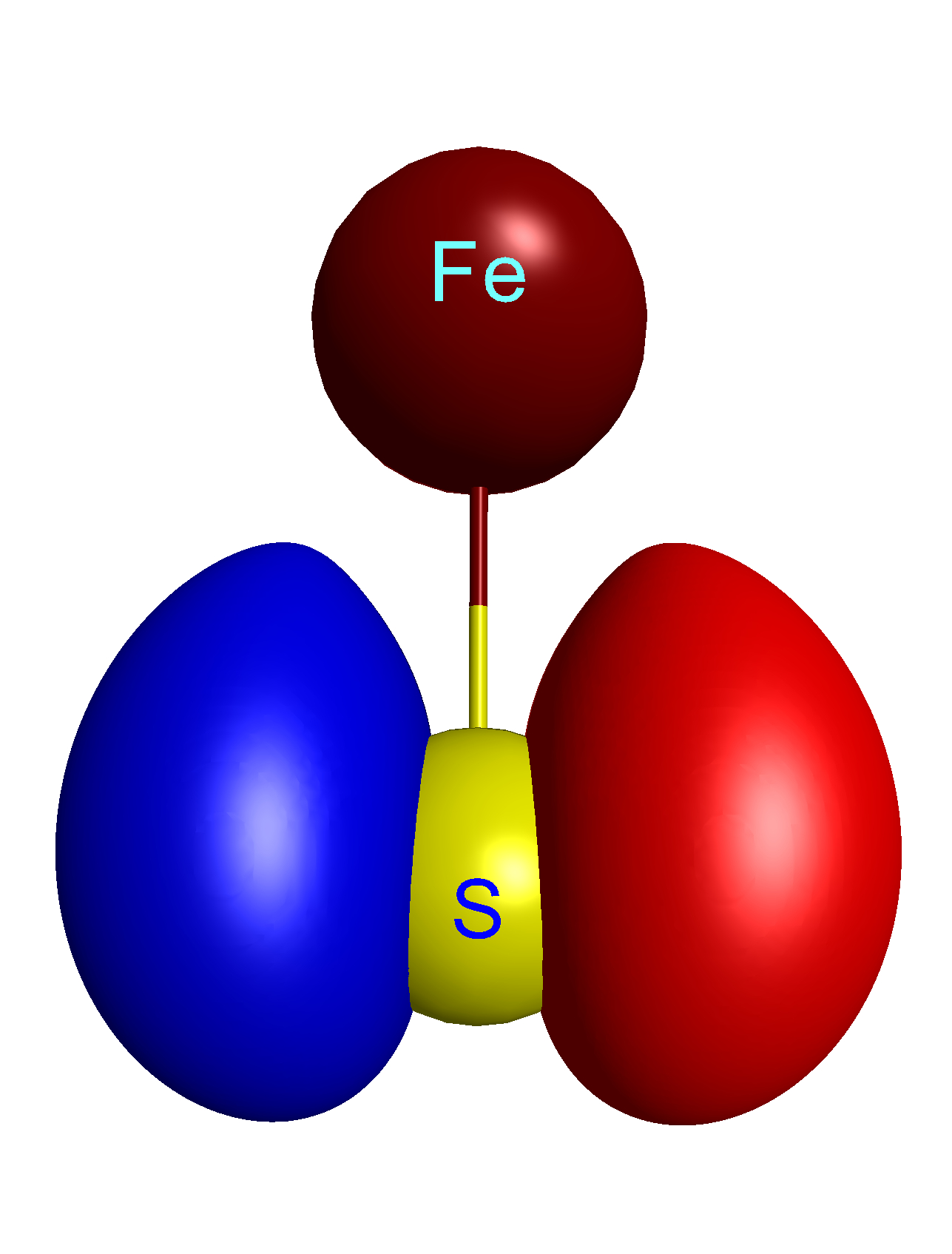}
     \end{minipage}
     \hspace{.00\linewidth}
     \begin{minipage}{.48\linewidth}
         \includegraphics[width=\linewidth]{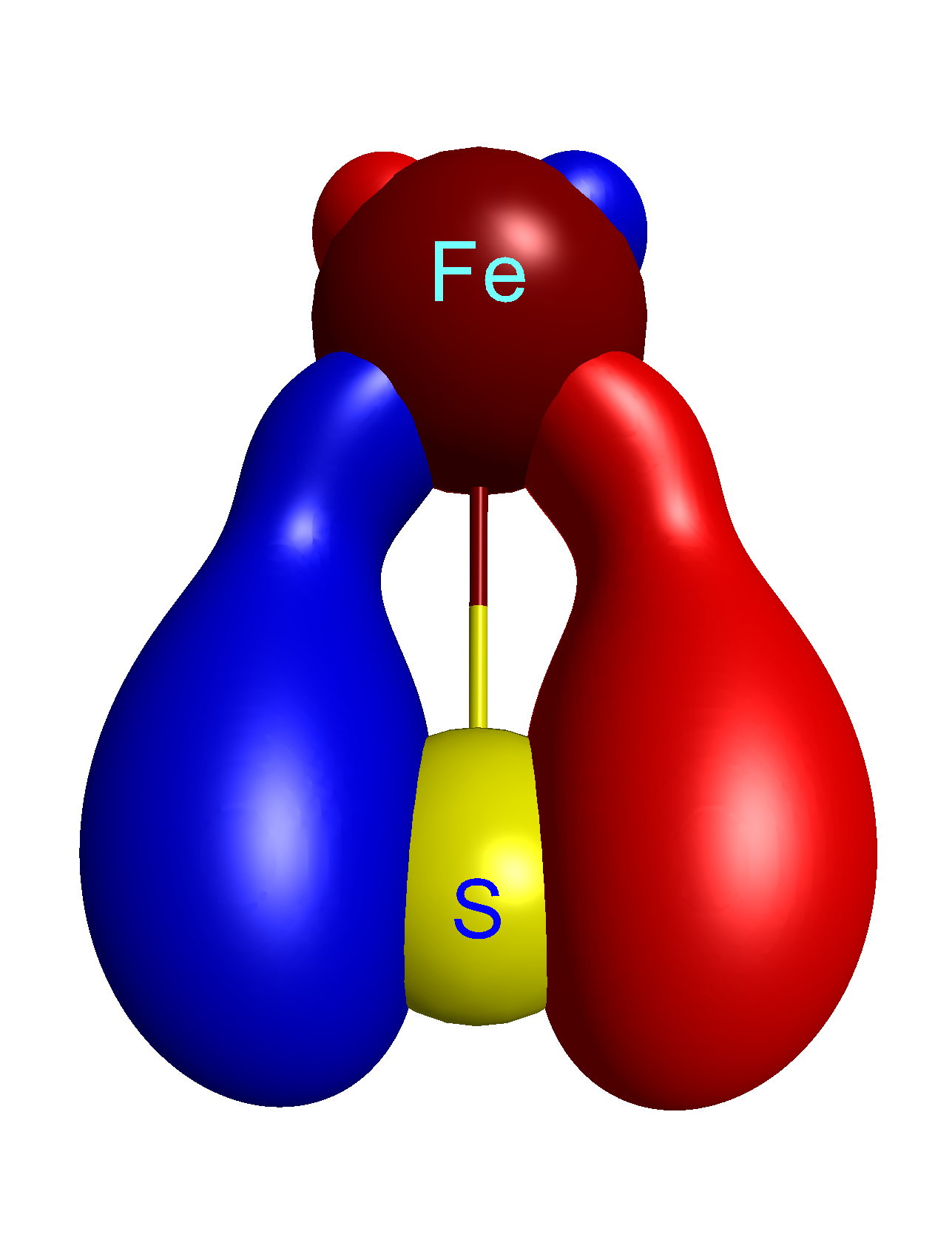}
     \end{minipage}
     \caption{$^5\Delta$  HF (left) and B3LYP (right) $\pi$ bonding orbitals.}
     \label{FIG:ORB}
 \end{figure}

The $^5\Delta$ energy is in all variants below the $^5\Sigma^+$ energy. The large difference
 between the DMC calculations with HF and B3LYP orbitals is very unusual. 
 A comparison of HF and B3LYP orbitals shows a significant difference only in the quadruply 
occupied $\pi$ orbital which has substantially more bonding character in B3LYP compared to HF, see Figure~\ref{FIG:ORB}.
  The same effect has been reported by Wagner and Mitas for some transition metal 
  oxides~\cite{wagner2003,Wagner2007m}. The B3LYP orbitals appear to be optimal because orbital optimization 
  in addition to Jastrow parameter optimization does not lower the DMC energy, the $^5\Sigma^+$ energy 
  is even slightly above the DMC energy with B3LYP orbitals. 

When comparing the DMC results with HF or B3LYP guide functions and both Jastrow factors one observes a reduction 
of the DMC energy by approximately 6~m$E_h$
when using the larger J2. Since the nodes of the guide function are unchanged this difference is due to the different localization 
of the pseudo potential. These results highlight the importance of using an accurate Jastrow factor for the localization of the 
pseudo potential. Note that J1 is not a small Jastrow factor as it already contains three-body electron-electron-nucleus terms.
When the orbitals are optimized in the presence of the Jastrow factor the more accurate J2 improves the nodes of the guide functions further in addition to reducing the pseudo potential localization error.

In Table~\ref{TAB:SingleConfVMC}, the corresponding VMC energies are shown. The last column containing the results for
energy-optimized orbitals (and simultaneous optimization of the Jastrow factor) demonstrates the effectiveness of the optimization.
Substantial systematic reduction of the VMC energy is obtained when using optimal orbitals instead of B3LYP orbitals 
or when using the larger Jastrow factor.

\begin{figure}
    \centering
    \begin{minipage}{.48\linewidth}
        \includegraphics[width=\linewidth]{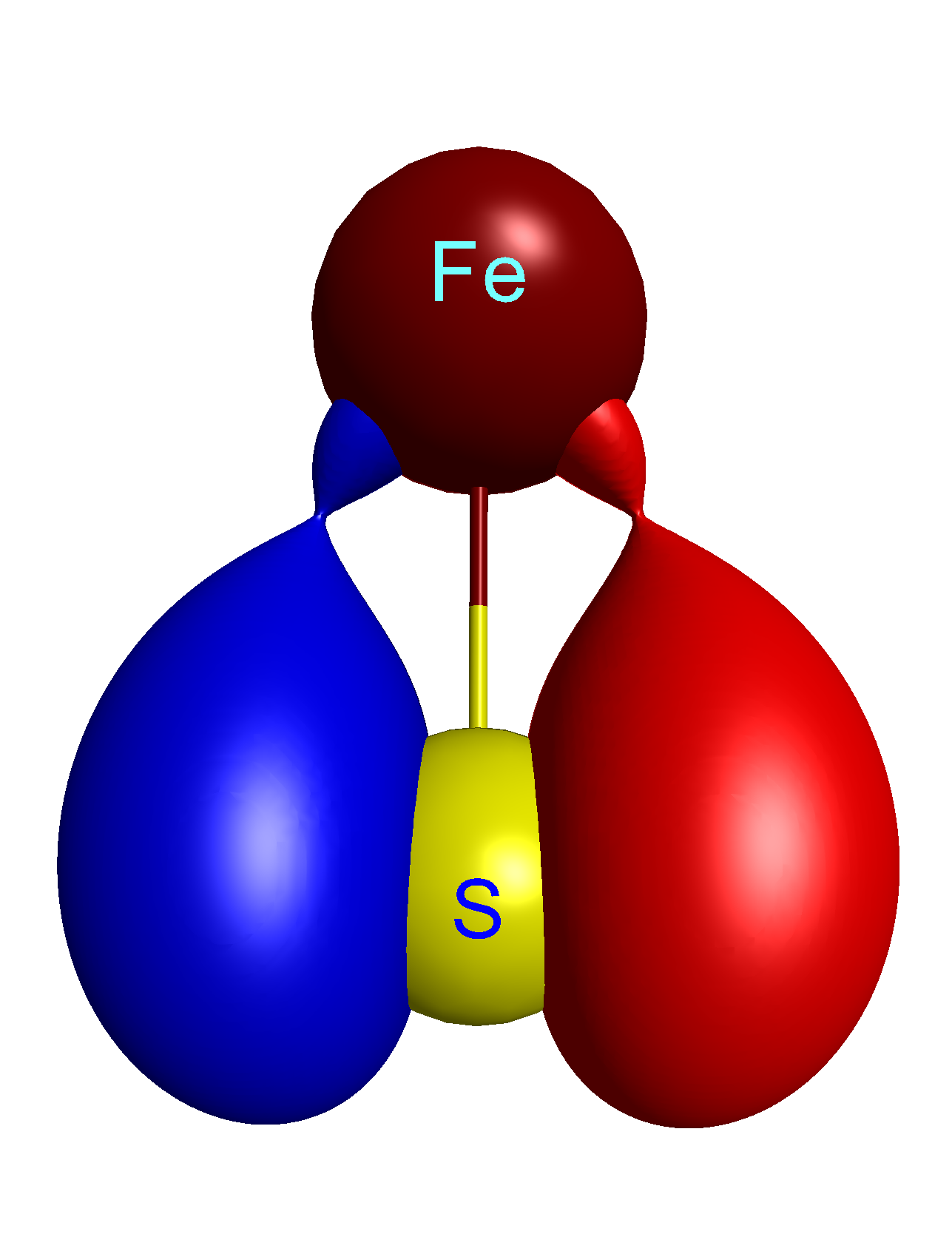}
    \end{minipage}
    \hspace{.00\linewidth}
    \begin{minipage}{.48\linewidth}
        \includegraphics[width=\linewidth]{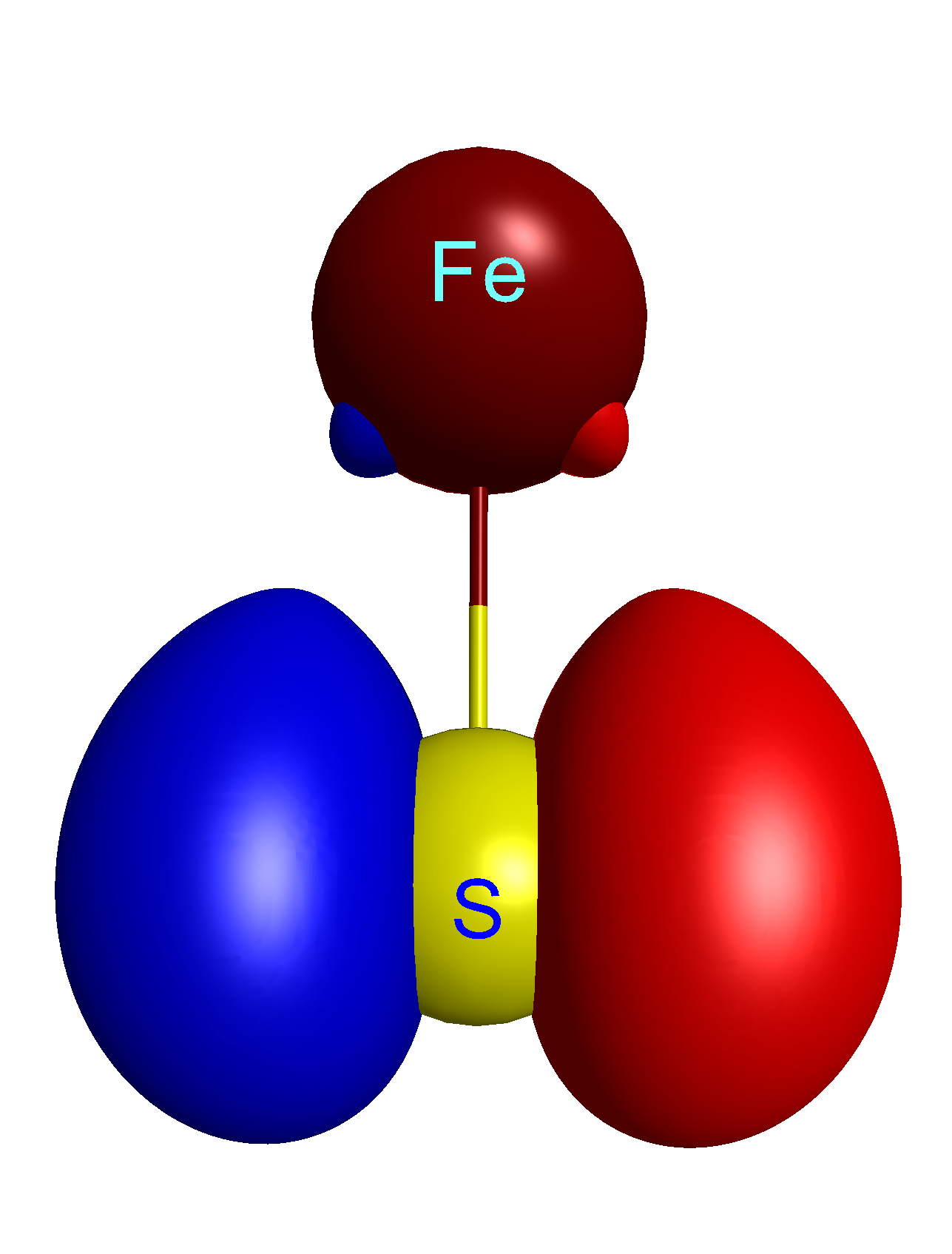}
    \end{minipage}
    \caption{$^5\Delta$  VMC optimized $\pi$ bonding orbitals with J1 (left) and J2 (right) Jastrow factors.}
    \label{FIG:ORBOPT}
\end{figure}

In Figure ~\ref{FIG:ORBOPT}, the quadruply occupied $\pi$ orbital is shown after energy optimization with both Jastrow
factors. Although the VMC energy is substantially lower compared to VMC with B3LYP orbitals, the $\pi$ orbitals display
much less bonding character and for the more accurate J2 even less than for J1.
We observe that the bonding character of the orbitals does not correlate with the stability of FeS. This fact may
hint to the importance of many-electron effects in transition metal bonding.

These calculations show a preference for 
the $^5\Delta$ state but do not result in a dissociation energy close to the experimental value because both states
 show significant multireference character and require a multideterminant guide function for accurate determination of
  the dissociation energy.

\subsection{Multideterminant QMC calculations}
The most common way to include nondynamical correlation is a CAS-type wave function. The CAS space described above results
 in 630 determinants for $^5\Sigma^+$ and 500 for $^5\Delta$. For efficient evaluation, the wave function is expressed 
 in terms of determinants while CSFs are employed for the parameter optimization. The Slater-Jastrow wave function is now 
 capable of covering nondynamical correlation with the determinant part and simultaneously dynamical correlation with the 
 Jastrow function. In Table~\ref{TAB:MDET}, DMC results are shown for both states with some or all parameters optimized 
 with respect to the energy. If not optimized, the CI and MO parameters are taken from the ab initio CASSCF wave function.
 The last column in Table~\ref{TAB:MDET} contains DMC energies with the fully optimized guide function employing the 
 J2 Jastrow factor.      
 All entries are a result of time step extrapolation. The DMC results are remarkable, in particular when comparing with the
  single determimant results in Table~\ref{TAB:SingleConf}.
For both states, all DMC energies are below the single determinant results with HF orbitals as expected, but surprisingly the
 single determinant results with B3LYP orbitals are substantially below the CAS-based DMC results even with optimized CI 
 coefficients. Only after optimization of the orbitals the energy is substantially improved compared with the DMC/B3LYP 
 result. 

After the discussion of the single determinant energies this result may not be fully surprising as the strong lowering of
 the DMC energy with B3LYP orbitals indicates a significant influence of the dynamical electron correlation on the 
 nodal structure. 
 This influence is missing when CASSCF orbitals are employed. Only the orbital optimization in the presence of a Jastrow 
 correlation function is capable of catching this influence. 
Optimization of the orbital parameters simultaneously with J1 Jastrow and CI parameters leads to an improvement of the energy
 by 14 and 8 m$E_h$ for the $^5\Delta$ and $^5\Sigma^+$ state, respectively, compared to optimization of J1 Jastrow and CI parameters only. 
 Note that the DMC energies with the CASSCF wave function show
  a $^5\Sigma^+$ ground state. Only after orbital optimization a $^5\Delta$ ground state is identified although the 
  energy difference is not statistically significant (as long as spin-orbit corrections are neglected, see below).  
  
This discussion allows us to understand why DFT calculates mainly a $^5\Delta$ ground state whereas high-level wave function
 methods yield a $^5\Sigma^+$ state.
DMC with CASSCF orbitals covers the dynamical correlation energy in a similar way (although more complete) as CASPT2 or ACPF. 
While the DMC accuracy is limited by the CASSCF nodes, CASPT2 and ACPF are limited by the excitation level out of the CASSCF reference
and the basis set.
When CAS orbitals optimized in the presence of a Jastrow factor (i.e. dynamic correlation) are used we observe a stabilization of 
$^5\Delta$ with respect to $^5\Sigma^+$. We expect the same result for CASPT2 if the optimization of the CAS orbitals for the full 
CASPT2 energy is done.

\begin{table}
		\caption{DMC energies in $E_h$ with CAS guide functions at different levels of parameter optimization as indicated. The Jastrow factor is indicated in parentheses.}
{\scriptsize 	\begin{tabular}{lllll}
		\hline\hline
		              &  Jas(J1)         & Jas(J1)+CI         & all(J1) & all(J2)  \\
		\hline
		$^5\Delta$    & -134.0245(4)  &  -134.0348(4)  &  -134.0489(3)&-134.0579(4)  \\
		$^5\Sigma^+$  & -134.0326(4)  &  -134.0406(3)  &  -134.0485(4)&-134.0571(4)  \\
		\hline\hline
	\end{tabular}}
\label{TAB:MDET}
\end{table}

\subsection{Calculation of the dissociation energy}

The electronic dissociation energy is obtained by subtracting the atomic DMC energies shown in Table ~\ref{TAB:Atom}.
There is only one CSF for each atom with the active space given above. The time step is extrapolated to zero.
In case of the atoms the changes of the DMC energy for different levels of optimization is not 
significant and is of the order of the statistical error. The comparison of the two Jastrow factors demonstrates once more the 
improved pseudopotential localization with the larger Jastrow factor.

\begin{table}
    \caption{Atomic DMC energies in $E_h$ with CAS guide function at different levels optimization and different Jastrow factors.}
    	\begin{tabular}{llll}
            \hline\hline
      &      &  Jas          & Jas+MO       \\
            \hline
S & J1       &  -10.1298(1)  & -10.1298(1)   \\
S & J2       &  -10.1309(1)  & -10.1314(1)  \\ 
Fe & J1      &  -123.8073(4) &  -123.8079(4) \\
Fe & J2      &  -123.8125(4) &  -123.8126(4) \\
            \hline\hline
    \end{tabular}
    \label{TAB:Atom}
    \end{table}

The experimental dissociation energy of $D_0 = 3.31(15)$ eV measured by Drowart et al. refers to 0 K~\cite{DrowartJ67}. For
 comparison with the experimental value the zero point energies (ZPE) and the spin-orbit contribution are required. Scalar
  relativistic effects are taken into account by the pseudo potentials. Finally, the core-valence correlation energy is 
  estimated.

Potential energy curves are calculated for both states at the DMC level with fully optimized multideterminant 
guide functions with the J1 Jastrow factor. For efficiency reasons, a fixed time step resulting in a 99\% acceptance ratio was 
employed. From a fit of the curves to Morse functions we obtain the spectroscopic constants and 
 the zero point energy shown in Table~\ref{TAB:ZPE}. 

The equilibrium bond length of 2.031(7) {\AA} is slightly longer than the experimental value 2.017 {\AA} obtained by 
microwave spectroscopy~\cite{Takano2004}. The harmonic frequencies and anharmonicities agree  with experimental 
values $\omega_e$ = 518(5) cm$^{-1}$ and $\omega_e x_e$ = 1.7(2) cm$^{-1}$ measured by Wang et al. ~\cite{Wang2011}. 
The zero point energy is slightly larger for the $^5\Sigma^+$ state.
 
\begin{table}
   \caption{Spectroscopic constants of FeS for fully optimized CAS wave function with Jastrow factor J1. Equilibrium distance is given in {\AA}, zero point energy in eV, and the harmonic frequency and anharmonicity in cm$^{-1}$.}
   \begin{tabular}{lllll}
      \hline\hline
      state        & $r_e$     & ZPE &$\omega_e$ & $\omega_e x_e$  \\
      \hline
      $^5\Sigma^+$ & 2.00(1)   & 0.0320(4) & 518(7)  & 2.67(7)   \\
      $^5\Delta$   & 2.031(7)  & 0.0308(7) & 499(11) & 2.53(11)  \\
      $^5\Delta$   & 2.017$^a$     & 0.0321(3)$^b$ & 518(5)$^b$  & 1.7(2)$^b$  \\ 
      \hline
     \end{tabular}
    \begin{tabular}{lllll}
      $^a$ derived from Takano et al.\cite{Takano2004} \hspace{1.75cm} \\
      $^b$ Wang et al.\cite{Wang2011}\\
          
      \hline\hline      
   \end{tabular}
   \label{TAB:ZPE}
\end{table}
                          
While spin-orbit effects cannot be included directly in the fixed node DMC framework, Melton at al.~\cite{Melton2016}
 were able to calculate spin-orbit interactions in DMC in fixed phase approximation. In this work, spin-orbit 
 corrections are taken from the literature. For the atoms, -0.024 eV for sulfur and -0.050 eV for iron are derived as first-order 
 spin-orbit corrections from experimental splittings~\cite{NIST_ASD}. The $^5\Sigma^+$ state shows no first-order 
 spin-orbit splitting while for the $^5\Delta$ state we obtain from Schultz et al.~\cite{Schultz2005a} a spin-orbit 
 correction of -0.022 eV. Thus, the spin-orbit correction reduces the dissociation energy of the $^5\Sigma^+$ state 
 by 0.074 eV and the dissociation energy of the $^5\Delta$  state by 0.052 eV. 

Due to the pseudopotentials, the valence correlation energy is calculated with QMC. Rather than attempting expensive 
all-electron QMC calculations, the core-valence correlation contribution is estimated with multireference perturbation 
theory known as MC-QDPT2 or MR-MP2 ~\cite{Nakano19931,Nakano1993} and implemented in GAMESS~\cite{Gamess}. While the same active space as in QMC is employed the core-valence basis set
 TK+NOSeC-V-QZP with all diffuse functions is used~\cite{Koga1999,Noro2000}. The dissociation energy is calculated with and without correlating the 
 core electrons to estimate the core-valence correlation contribution to the dissociation energy. 
 This contribution is surprisingly large. For the $^5\Delta$ state, the dissociation energy increases by the core-valence
  correlation contribution by about 0.143 eV and by about 0.146 eV for the $^5\Sigma^+$ state.

With the DMC energies from fully optimized wave functions using the J2 Jastrow factor and all of the above corrections 
the energy gap between $^5\Delta$ and $^5\Sigma^+$ increases to 
0.042(15) eV, confirming the $^5\Delta$ state as 
ground state in agreement with experiment.

Adding all corrections to the electronic dissociation energy for the $^5\Delta$ state we obtain +0.060 eV. Table ~\ref{TAB:De} shows
 the DMC dissociation energy $D_0$ for the different guide wave functions with this correction added. 
Moderate agreement with the experimental value is obtained when unrestricted B3LYP orbitals are employed for the atoms and the 
molecule. The agreement deteriorates with both Jastrow factors slightly when using restricted determinants with optimized orbitals 
(where the symmetry of the orbitals is preserved during the optimization). Substantially stronger deviation from the experimental
value is obtained not only with an unoptimized CASSCF wave function where only the Jastrow part is optimized, but also when
optimizing the CI coefficients of the CAS wave function. Finally, improvement over the unrestricted single determinant energy 
is calculated with full optimization of the CAS wave function, i.e. optimization of orbitals, CI coefficients, and 
Jastrow parameters simultaneously. With the large J2 Jastrow function, excellent agreement with the experimental value is obtained.
The significant effect of the Jastrow factor on the DMC dissociation energy was surprising to us. As discussed above it is due
to the pseudo potential localization and the effect of the Jastrow factor on the orbital optimization. The additional CPU cost for 
the substantially larger J2 Jastrow factor is, on average, only 10\%.

	\begin{table}
		\caption{DMC dissociation energy $D_0$ in eV for FeS ($^5\Delta$) with different guide functions and different optimized parameters (opt params). Spin-orbit, zero
             point energy, and core valence correlation corrections are included.}
	\begin{center}
		
		\begin{tabular}{l l l l l }
			\hline\hline
			nodes    & Jas  & opt params    & $D_0$       \\ \hline
			UB3LYP   & J1  & Jas         &  2.988(16)    \\
            1 Det    & J1  & Jas+MO      &  2.913(16)    \\
            1 Det    & J2  & Jas+MO      &  2.914(15)    \\
			CASSCF   & J1  & Jas         &  2.459(16)    \\
			CAS      & J1  & Jas+CI      &  2.720(16)    \\
			CAS      & J1  & Jas+MO+CI   &  3.085(15)    \\
            CAS      & J2  & Jas+MO+CI   &  3.159(15)    \\
			exp      &     &             &  3.31(15)$^a$ \\
         \hline
         $^a$ Drowart et al.\cite{DrowartJ67}\\
         \hline\hline
		\end{tabular}
	\end{center}
\label{TAB:De}
\end{table}

\subsection{Conclusion}
Single and CAS-based multideterminant DMC calculations have been carried out with partial and full optimization of
Jastrow, CI, and orbital parameters.
The $^5\Delta$ state was identified as ground state, in agreement with experiment but in disagreement with other
wave function methods.
$^5\Delta$ is obtained as ground state only after optimization of the CAS orbitals simultaneously with the Jastrow and CI parameters. 
The CASSCF reference seems to be biased toward the $^5\Sigma+$ state which can explain why CASPT2 yields a $^5\Sigma^+$
ground state.
The optimization of the CAS orbitals in the presence of the Jastrow correlation function allows thus to construct
compact accurate wave functions yielding excellent energies even for challenging molecules such as FeS. 
In addition to the dissociation energy, a potential curve was calculated, yielding harmonic frequencies and anharmonicities 
in good agreement with experiment.
Furthermore,
the influence of the Jastrow factor on the DMC results has been investigated. While most DMC calculations on transition metal compounds employed 
small Jastrow factors, the above results indicate that a large Jastrow factor allows for higher DMC accuracy through improved
pseudo potential localization and improved optimized orbitals. The additional cost for a larger Jastrow factor is small in
pseudo potential calculations.

\begin{acknowledgments}
The authors gratefully acknowledge computing resources granted by J\"ulich-Aachen research alliance (JARA) and RWTH Aachen 
University under projects jara0119 and rwth0128.
\end{acknowledgments}
\bibliographystyle{apsrev4-1} 
\bibliography{bib} 

\begin{thebibliography}{49}%
\makeatletter
\providecommand \@ifxundefined [1]{%
 \@ifx{#1\undefined}
}%
\providecommand \@ifnum [1]{%
 \ifnum #1\expandafter \@firstoftwo
 \else \expandafter \@secondoftwo
 \fi
}%
\providecommand \@ifx [1]{%
 \ifx #1\expandafter \@firstoftwo
 \else \expandafter \@secondoftwo
 \fi
}%
\providecommand \natexlab [1]{#1}%
\providecommand \enquote  [1]{``#1''}%
\providecommand \bibnamefont  [1]{#1}%
\providecommand \bibfnamefont [1]{#1}%
\providecommand \citenamefont [1]{#1}%
\providecommand \href@noop [0]{\@secondoftwo}%
\providecommand \href [0]{\begingroup \@sanitize@url \@href}%
\providecommand \@href[1]{\@@startlink{#1}\@@href}%
\providecommand \@@href[1]{\endgroup#1\@@endlink}%
\providecommand \@sanitize@url [0]{\catcode `\\12\catcode `\$12\catcode
  `\&12\catcode `\#12\catcode `\^12\catcode `\_12\catcode `\%12\relax}%
\providecommand \@@startlink[1]{}%
\providecommand \@@endlink[0]{}%
\providecommand \url  [0]{\begingroup\@sanitize@url \@url }%
\providecommand \@url [1]{\endgroup\@href {#1}{\urlprefix }}%
\providecommand \urlprefix  [0]{URL }%
\providecommand \Eprint [0]{\href }%
\providecommand \doibase [0]{http://dx.doi.org/}%
\providecommand \selectlanguage [0]{\@gobble}%
\providecommand \bibinfo  [0]{\@secondoftwo}%
\providecommand \bibfield  [0]{\@secondoftwo}%
\providecommand \translation [1]{[#1]}%
\providecommand \BibitemOpen [0]{}%
\providecommand \bibitemStop [0]{}%
\providecommand \bibitemNoStop [0]{.\EOS\space}%
\providecommand \EOS [0]{\spacefactor3000\relax}%
\providecommand \BibitemShut  [1]{\csname bibitem#1\endcsname}%
\let\auto@bib@innerbib\@empty
\bibitem [{\citenamefont {Wagner}(2014)}]{Wagner2014}%
  \BibitemOpen
  \bibfield  {author} {\bibinfo {author} {\bibfnamefont {L.~K.}\ \bibnamefont
  {Wagner}},\ }\href {\doibase 10.1002/qua.24526} {\bibfield  {journal}
  {\bibinfo  {journal} {Int. J. Quantum Chem.}\ }\textbf {\bibinfo {volume}
  {114}},\ \bibinfo {pages} {94} (\bibinfo {year} {2014})}\BibitemShut
  {NoStop}%
\bibitem [{\citenamefont {Austin}\ \emph {et~al.}(2012)\citenamefont {Austin},
  \citenamefont {Zubarev},\ and\ \citenamefont {Lester}}]{Austin2012}%
  \BibitemOpen
  \bibfield  {author} {\bibinfo {author} {\bibfnamefont {B.~M.}\ \bibnamefont
  {Austin}}, \bibinfo {author} {\bibfnamefont {D.~Y.}\ \bibnamefont {Zubarev}},
  \ and\ \bibinfo {author} {\bibfnamefont {W.~A.}\ \bibnamefont {Lester}},\
  }\href {\doibase 10.1021/cr2001564} {\bibfield  {journal} {\bibinfo
  {journal} {Chem. Rev.}\ }\textbf {\bibinfo {volume} {112}},\ \bibinfo {pages}
  {263} (\bibinfo {year} {2012})}\BibitemShut {NoStop}%
\bibitem [{\citenamefont {L{\"{u}}chow}(2011)}]{Luechow2011}%
  \BibitemOpen
  \bibfield  {author} {\bibinfo {author} {\bibfnamefont {A.}~\bibnamefont
  {L{\"{u}}chow}},\ }\href {\doibase 10.1002/wcms.40} {\bibfield  {journal}
  {\bibinfo  {journal} {WIREs Comput. Mol. Sci.}\ }\textbf {\bibinfo {volume}
  {1}},\ \bibinfo {pages} {388} (\bibinfo {year} {2011})}\BibitemShut {NoStop}%
\bibitem [{\citenamefont {Grossman}(2002)}]{grossman2002}%
  \BibitemOpen
  \bibfield  {author} {\bibinfo {author} {\bibfnamefont {J.~C.}\ \bibnamefont
  {Grossman}},\ }\href {\doibase 10.1063/1.1487829} {\bibfield  {journal}
  {\bibinfo  {journal} {J. Chem. Phys.}\ }\textbf {\bibinfo {volume} {117}},\
  \bibinfo {pages} {1434} (\bibinfo {year} {2002})}\BibitemShut {NoStop}%
\bibitem [{\citenamefont {Nemec}\ \emph {et~al.}(2010)\citenamefont {Nemec},
  \citenamefont {Towler},\ and\ \citenamefont {Needs}}]{Nemec2010}%
  \BibitemOpen
  \bibfield  {author} {\bibinfo {author} {\bibfnamefont {N.}~\bibnamefont
  {Nemec}}, \bibinfo {author} {\bibfnamefont {M.~D.}\ \bibnamefont {Towler}}, \
  and\ \bibinfo {author} {\bibfnamefont {R.~J.}\ \bibnamefont {Needs}},\ }\href
  {\doibase 10.1063/1.3288054} {\bibfield  {journal} {\bibinfo  {journal} {J.
  Chem. Phys.}\ }\textbf {\bibinfo {volume} {132}},\ \bibinfo {pages} {034111}
  (\bibinfo {year} {2010})}\BibitemShut {NoStop}%
\bibitem [{\citenamefont {Kannengie{\ss}er}\ \emph {et~al.}(2014)\citenamefont
  {Kannengie{\ss}er}, \citenamefont {Klahm}, \citenamefont {{Vinh Lam Nguyen}},
  \citenamefont {L{\"{u}}chow},\ and\ \citenamefont {Stahl}}]{klahm20141}%
  \BibitemOpen
  \bibfield  {author} {\bibinfo {author} {\bibfnamefont {R.}~\bibnamefont
  {Kannengie{\ss}er}}, \bibinfo {author} {\bibfnamefont {S.}~\bibnamefont
  {Klahm}}, \bibinfo {author} {\bibfnamefont {H.}~\bibnamefont {{Vinh Lam
  Nguyen}}}, \bibinfo {author} {\bibfnamefont {A.}~\bibnamefont
  {L{\"{u}}chow}}, \ and\ \bibinfo {author} {\bibfnamefont {W.}~\bibnamefont
  {Stahl}},\ }\href {\doibase 10.1063/1.4901980} {\bibfield  {journal}
  {\bibinfo  {journal} {J. Chem. Phys.}\ }\textbf {\bibinfo {volume} {141}},\
  \bibinfo {pages} {204308} (\bibinfo {year} {2014})}\BibitemShut {NoStop}%
\bibitem [{\citenamefont {Klahm}\ and\ \citenamefont
  {L{\"{u}}chow}(2014)}]{Klahm20147}%
  \BibitemOpen
  \bibfield  {author} {\bibinfo {author} {\bibfnamefont {S.}~\bibnamefont
  {Klahm}}\ and\ \bibinfo {author} {\bibfnamefont {A.}~\bibnamefont
  {L{\"{u}}chow}},\ }\href {\doibase 10.1016/j.cplett.2014.03.044} {\bibfield
  {journal} {\bibinfo  {journal} {Chem. Phys. Lett.}\ }\textbf {\bibinfo
  {volume} {600}},\ \bibinfo {pages} {7} (\bibinfo {year} {2014})}\BibitemShut
  {NoStop}%
\bibitem [{\citenamefont {Dubeck{\'{y}}}\ \emph {et~al.}(2016)\citenamefont
  {Dubeck{\'{y}}}, \citenamefont {Mitas},\ and\ \citenamefont
  {Jure{\v{c}}ka}}]{Dubecky2016}%
  \BibitemOpen
  \bibfield  {author} {\bibinfo {author} {\bibfnamefont {M.}~\bibnamefont
  {Dubeck{\'{y}}}}, \bibinfo {author} {\bibfnamefont {L.}~\bibnamefont
  {Mitas}}, \ and\ \bibinfo {author} {\bibfnamefont {P.}~\bibnamefont
  {Jure{\v{c}}ka}},\ }\href {\doibase 10.1021/acs.chemrev.5b00577} {\bibfield
  {journal} {\bibinfo  {journal} {Chem. Rev.}\ }\textbf {\bibinfo {volume}
  {116}},\ \bibinfo {pages} {5188} (\bibinfo {year} {2016})}\BibitemShut
  {NoStop}%
\bibitem [{\citenamefont {Sokolova}\ and\ \citenamefont
  {L{\"{u}}chow}(2000)}]{Sokolova2000}%
  \BibitemOpen
  \bibfield  {author} {\bibinfo {author} {\bibfnamefont {S.}~\bibnamefont
  {Sokolova}}\ and\ \bibinfo {author} {\bibfnamefont {A.}~\bibnamefont
  {L{\"{u}}chow}},\ }\href {\doibase 10.1016/S0009-2614(00)00276-1} {\bibfield
  {journal} {\bibinfo  {journal} {Chem. Phys. Lett.}\ }\textbf {\bibinfo
  {volume} {320}},\ \bibinfo {pages} {421} (\bibinfo {year}
  {2000})}\BibitemShut {NoStop}%
\bibitem [{\citenamefont {Wagner}\ and\ \citenamefont
  {Mitas}(2007)}]{Wagner2007m}%
  \BibitemOpen
  \bibfield  {author} {\bibinfo {author} {\bibfnamefont {L.~K.}\ \bibnamefont
  {Wagner}}\ and\ \bibinfo {author} {\bibfnamefont {L.}~\bibnamefont {Mitas}},\
  }\href {\doibase 10.1063/1.2428294} {\bibfield  {journal} {\bibinfo
  {journal} {J. Chem. Phys.}\ }\textbf {\bibinfo {volume} {126}},\ \bibinfo
  {pages} {034105} (\bibinfo {year} {2007})}\BibitemShut {NoStop}%
\bibitem [{\citenamefont {Wagner}\ and\ \citenamefont
  {Mitas}(2003)}]{wagner2003}%
  \BibitemOpen
  \bibfield  {author} {\bibinfo {author} {\bibfnamefont {L.}~\bibnamefont
  {Wagner}}\ and\ \bibinfo {author} {\bibfnamefont {L.}~\bibnamefont {Mitas}},\
  }\href {\doibase 10.1016/S0009-2614(03)00128-3} {\bibfield  {journal}
  {\bibinfo  {journal} {Chem. Phys. Lett.}\ }\textbf {\bibinfo {volume}
  {370}},\ \bibinfo {pages} {412} (\bibinfo {year} {2003})}\BibitemShut
  {NoStop}%
\bibitem [{\citenamefont {Diedrich}\ \emph {et~al.}(2005)\citenamefont
  {Diedrich}, \citenamefont {L{\"{u}}chow},\ and\ \citenamefont
  {Grimme}}]{Diedrich2005}%
  \BibitemOpen
  \bibfield  {author} {\bibinfo {author} {\bibfnamefont {C.}~\bibnamefont
  {Diedrich}}, \bibinfo {author} {\bibfnamefont {A.}~\bibnamefont
  {L{\"{u}}chow}}, \ and\ \bibinfo {author} {\bibfnamefont {S.}~\bibnamefont
  {Grimme}},\ }\href {\doibase 10.1063/1.1846654} {\bibfield  {journal}
  {\bibinfo  {journal} {J. Chem. Phys.}\ }\textbf {\bibinfo {volume} {122}},\
  \bibinfo {pages} {021101} (\bibinfo {year} {2005})}\BibitemShut {NoStop}%
\bibitem [{\citenamefont {Bande}\ and\ \citenamefont
  {L{\"{u}}chow}(2008)}]{Bande2008}%
  \BibitemOpen
  \bibfield  {author} {\bibinfo {author} {\bibfnamefont {A.}~\bibnamefont
  {Bande}}\ and\ \bibinfo {author} {\bibfnamefont {A.}~\bibnamefont
  {L{\"{u}}chow}},\ }\href {\doibase 10.1039/b803571g} {\bibfield  {journal}
  {\bibinfo  {journal} {Phys. Chem. Chem. Phys.}\ }\textbf {\bibinfo {volume}
  {10}},\ \bibinfo {pages} {3371} (\bibinfo {year} {2008})}\BibitemShut
  {NoStop}%
\bibitem [{\citenamefont {Doblhoff-Dier}\ \emph {et~al.}(2016)\citenamefont
  {Doblhoff-Dier}, \citenamefont {Meyer}, \citenamefont {Hoggan}, \citenamefont
  {Kroes},\ and\ \citenamefont {Wagner}}]{Doblhoff-Dier2016}%
  \BibitemOpen
  \bibfield  {author} {\bibinfo {author} {\bibfnamefont {K.}~\bibnamefont
  {Doblhoff-Dier}}, \bibinfo {author} {\bibfnamefont {J.}~\bibnamefont
  {Meyer}}, \bibinfo {author} {\bibfnamefont {P.~E.}\ \bibnamefont {Hoggan}},
  \bibinfo {author} {\bibfnamefont {G.-J.}\ \bibnamefont {Kroes}}, \ and\
  \bibinfo {author} {\bibfnamefont {L.~K.}\ \bibnamefont {Wagner}},\ }\href
  {\doibase 10.1021/acs.jctc.6b00160} {\bibfield  {journal} {\bibinfo
  {journal} {J. Chem. Theory Comput.}\ }\textbf {\bibinfo {volume} {12}},\
  \bibinfo {pages} {2583} (\bibinfo {year} {2016})}\BibitemShut {NoStop}%
\bibitem [{\citenamefont {Casula}\ \emph {et~al.}(2009)\citenamefont {Casula},
  \citenamefont {Marchi}, \citenamefont {Azadi},\ and\ \citenamefont
  {Sorella}}]{Casula2009}%
  \BibitemOpen
  \bibfield  {author} {\bibinfo {author} {\bibfnamefont {M.}~\bibnamefont
  {Casula}}, \bibinfo {author} {\bibfnamefont {M.}~\bibnamefont {Marchi}},
  \bibinfo {author} {\bibfnamefont {S.}~\bibnamefont {Azadi}}, \ and\ \bibinfo
  {author} {\bibfnamefont {S.}~\bibnamefont {Sorella}},\ }\href {\doibase
  10.1016/j.cplett.2009.07.005} {\bibfield  {journal} {\bibinfo  {journal}
  {Chem. Phys. Lett.}\ }\textbf {\bibinfo {volume} {477}},\ \bibinfo {pages}
  {255} (\bibinfo {year} {2009})}\BibitemShut {NoStop}%
\bibitem [{\citenamefont {Caffarel}\ \emph {et~al.}(2014)\citenamefont
  {Caffarel}, \citenamefont {Giner}, \citenamefont {Scemama},\ and\
  \citenamefont {Ram{\'{i}}rez-Sol{\'{i}}s}}]{Caffarel2014}%
  \BibitemOpen
  \bibfield  {author} {\bibinfo {author} {\bibfnamefont {M.}~\bibnamefont
  {Caffarel}}, \bibinfo {author} {\bibfnamefont {E.}~\bibnamefont {Giner}},
  \bibinfo {author} {\bibfnamefont {A.}~\bibnamefont {Scemama}}, \ and\
  \bibinfo {author} {\bibfnamefont {A.}~\bibnamefont
  {Ram{\'{i}}rez-Sol{\'{i}}s}},\ }\href {\doibase 10.1021/ct5004252} {\bibfield
   {journal} {\bibinfo  {journal} {J. Chem. Theory Comput.}\ }\textbf {\bibinfo
  {volume} {10}},\ \bibinfo {pages} {5286} (\bibinfo {year}
  {2014})}\BibitemShut {NoStop}%
\bibitem [{\citenamefont {Petz}\ and\ \citenamefont
  {L{\"{u}}chow}(2011)}]{Petz2011}%
  \BibitemOpen
  \bibfield  {author} {\bibinfo {author} {\bibfnamefont {R.}~\bibnamefont
  {Petz}}\ and\ \bibinfo {author} {\bibfnamefont {A.}~\bibnamefont
  {L{\"{u}}chow}},\ }\href {\doibase 10.1002/cphc.201000942} {\bibfield
  {journal} {\bibinfo  {journal} {ChemPhysChem}\ }\textbf {\bibinfo {volume}
  {12}},\ \bibinfo {pages} {2031} (\bibinfo {year} {2011})}\BibitemShut
  {NoStop}%
\bibitem [{\citenamefont {Hammond}\ \emph {et~al.}(1994)\citenamefont
  {Hammond}, \citenamefont {Lester},\ and\ \citenamefont
  {Reynolds.}}]{HammondQMC}%
  \BibitemOpen
  \bibfield  {author} {\bibinfo {author} {\bibfnamefont {B.~L.}\ \bibnamefont
  {Hammond}}, \bibinfo {author} {\bibfnamefont {W.~A.}\ \bibnamefont {Lester}},
  \ and\ \bibinfo {author} {\bibfnamefont {P.~J.}\ \bibnamefont {Reynolds.}},\
  }\href@noop {} {\emph {\bibinfo {title} {{Monte Carlo Methods in Ab Initio
  Quantum Chemistry}}}}\ (\bibinfo  {publisher} {World Scientific},\ \bibinfo
  {address} {Singapore},\ \bibinfo {year} {1994})\BibitemShut {NoStop}%
\bibitem [{\citenamefont {Reynolds}\ \emph {et~al.}(1982)\citenamefont
  {Reynolds}, \citenamefont {Ceperley}, \citenamefont {Alder},\ and\
  \citenamefont {Lester}}]{Reynolds82}%
  \BibitemOpen
  \bibfield  {author} {\bibinfo {author} {\bibfnamefont {P.~J.}\ \bibnamefont
  {Reynolds}}, \bibinfo {author} {\bibfnamefont {D.~M.}\ \bibnamefont
  {Ceperley}}, \bibinfo {author} {\bibfnamefont {B.~J.}\ \bibnamefont {Alder}},
  \ and\ \bibinfo {author} {\bibfnamefont {W.~A.}\ \bibnamefont {Lester}},\
  }\href {\doibase 10.1063/1.443766} {\bibfield  {journal} {\bibinfo  {journal}
  {J. Chem. Phys.}\ }\textbf {\bibinfo {volume} {77}},\ \bibinfo {pages} {5593}
  (\bibinfo {year} {1982})}\BibitemShut {NoStop}%
\bibitem [{\citenamefont {Schmidt}\ and\ \citenamefont
  {Moskowitz}(1990)}]{Schmidt1990}%
  \BibitemOpen
  \bibfield  {author} {\bibinfo {author} {\bibfnamefont {K.~E.}\ \bibnamefont
  {Schmidt}}\ and\ \bibinfo {author} {\bibfnamefont {J.~W.}\ \bibnamefont
  {Moskowitz}},\ }\href {\doibase 10.1063/1.458750} {\bibfield  {journal}
  {\bibinfo  {journal} {J. Chem Phy.}\ }\textbf {\bibinfo {volume} {93}},\
  \bibinfo {pages} {4172} (\bibinfo {year} {1990})}\BibitemShut {NoStop}%
\bibitem [{\citenamefont {L{\"{u}}chow}\ \emph {et~al.}(2015)\citenamefont
  {L{\"{u}}chow}, \citenamefont {Sturm}, \citenamefont {Schulte},\ and\
  \citenamefont {{Haghighi Mood}}}]{Luchow2015}%
  \BibitemOpen
  \bibfield  {author} {\bibinfo {author} {\bibfnamefont {A.}~\bibnamefont
  {L{\"{u}}chow}}, \bibinfo {author} {\bibfnamefont {A.}~\bibnamefont {Sturm}},
  \bibinfo {author} {\bibfnamefont {C.}~\bibnamefont {Schulte}}, \ and\
  \bibinfo {author} {\bibfnamefont {K.}~\bibnamefont {{Haghighi Mood}}},\
  }\href {\doibase 10.1063/1.4909554} {\bibfield  {journal} {\bibinfo
  {journal} {J. Chem. Phys.}\ }\textbf {\bibinfo {volume} {142}},\ \bibinfo
  {pages} {084111} (\bibinfo {year} {2015})}\BibitemShut {NoStop}%
\bibitem [{\citenamefont {Burkatzki}\ \emph {et~al.}(2008)\citenamefont
  {Burkatzki}, \citenamefont {Filippi},\ and\ \citenamefont
  {Dolg}}]{Burkatzki2008}%
  \BibitemOpen
  \bibfield  {author} {\bibinfo {author} {\bibfnamefont {M.}~\bibnamefont
  {Burkatzki}}, \bibinfo {author} {\bibfnamefont {C.}~\bibnamefont {Filippi}},
  \ and\ \bibinfo {author} {\bibfnamefont {M.}~\bibnamefont {Dolg}},\ }\href
  {\doibase 10.1063/1.2987872} {\bibfield  {journal} {\bibinfo  {journal} {J.
  Chem. Phys.}\ }\textbf {\bibinfo {volume} {129}},\ \bibinfo {pages} {164115}
  (\bibinfo {year} {2008})}\BibitemShut {NoStop}%
\bibitem [{\citenamefont {Burkatzki}\ \emph {et~al.}(2007)\citenamefont
  {Burkatzki}, \citenamefont {Filippi},\ and\ \citenamefont
  {Dolg}}]{burkatzki2007-234105}%
  \BibitemOpen
  \bibfield  {author} {\bibinfo {author} {\bibfnamefont {M.}~\bibnamefont
  {Burkatzki}}, \bibinfo {author} {\bibfnamefont {C.}~\bibnamefont {Filippi}},
  \ and\ \bibinfo {author} {\bibfnamefont {M.}~\bibnamefont {Dolg}},\ }\href
  {\doibase 10.1063/1.2741534} {\bibfield  {journal} {\bibinfo  {journal} {J.
  Chem. Phys.}\ }\textbf {\bibinfo {volume} {126}},\ \bibinfo {pages} {234105}
  (\bibinfo {year} {2007})}\BibitemShut {NoStop}%
\bibitem [{\citenamefont {Mitáš}\ \emph {et~al.}(1991)\citenamefont
  {Mitáš}, \citenamefont {Shirley},\ and\ \citenamefont
  {Ceperley}}]{Mitas1991}%
  \BibitemOpen
  \bibfield  {author} {\bibinfo {author} {\bibfnamefont {L.}~\bibnamefont
  {Mitáš}}, \bibinfo {author} {\bibfnamefont {E.~L.}\ \bibnamefont
  {Shirley}}, \ and\ \bibinfo {author} {\bibfnamefont {D.~M.}\ \bibnamefont
  {Ceperley}},\ }\href {\doibase 10.1063/1.460849} {\bibfield  {journal}
  {\bibinfo  {journal} {J. Chem. Phys.}\ }\textbf {\bibinfo {volume} {95}},\
  \bibinfo {pages} {3467} (\bibinfo {year} {1991})}\BibitemShut {NoStop}%
\bibitem [{\citenamefont {Toulouse}\ and\ \citenamefont
  {Umrigar}(2007)}]{Toulouse2007}%
  \BibitemOpen
  \bibfield  {author} {\bibinfo {author} {\bibfnamefont {J.}~\bibnamefont
  {Toulouse}}\ and\ \bibinfo {author} {\bibfnamefont {C.~J.}\ \bibnamefont
  {Umrigar}},\ }\href {\doibase 10.1063/1.2437215} {\bibfield  {journal}
  {\bibinfo  {journal} {J. Chem. Phys.}\ }\textbf {\bibinfo {volume} {126}},\
  \bibinfo {pages} {084102} (\bibinfo {year} {2007})}\BibitemShut {NoStop}%
\bibitem [{\citenamefont {Manten}\ and\ \citenamefont
  {L{\"{u}}chow}(2001)}]{Manten_JCP115_5362}%
  \BibitemOpen
  \bibfield  {author} {\bibinfo {author} {\bibfnamefont {S.}~\bibnamefont
  {Manten}}\ and\ \bibinfo {author} {\bibfnamefont {A.}~\bibnamefont
  {L{\"{u}}chow}},\ }\href {\doibase 10.1063/1.1394757} {\bibfield  {journal}
  {\bibinfo  {journal} {J. Chem. Phys.}\ }\textbf {\bibinfo {volume} {115}},\
  \bibinfo {pages} {5362} (\bibinfo {year} {2001})}\BibitemShut {NoStop}%
\bibitem [{\citenamefont {Scemama}\ \emph {et~al.}(2016)\citenamefont
  {Scemama}, \citenamefont {Applencourt}, \citenamefont {Giner},\ and\
  \citenamefont {Caffarel}}]{Scemama2016}%
  \BibitemOpen
  \bibfield  {author} {\bibinfo {author} {\bibfnamefont {A.}~\bibnamefont
  {Scemama}}, \bibinfo {author} {\bibfnamefont {T.}~\bibnamefont
  {Applencourt}}, \bibinfo {author} {\bibfnamefont {E.}~\bibnamefont {Giner}},
  \ and\ \bibinfo {author} {\bibfnamefont {M.}~\bibnamefont {Caffarel}},\
  }\href {\doibase 10.1002/jcc.24382} {\bibfield  {journal} {\bibinfo
  {journal} {J. Comput. Chem.}\ }\textbf {\bibinfo {volume} {37}},\ \bibinfo
  {pages} {1866} (\bibinfo {year} {2016})}\BibitemShut {NoStop}%
\bibitem [{\citenamefont {Zhang}\ \emph {et~al.}(1996)\citenamefont {Zhang},
  \citenamefont {Hayase}, \citenamefont {Kawamata}, \citenamefont {Nakao},
  \citenamefont {Nakajima},\ and\ \citenamefont {Kaya}}]{Zhang1996}%
  \BibitemOpen
  \bibfield  {author} {\bibinfo {author} {\bibfnamefont {N.}~\bibnamefont
  {Zhang}}, \bibinfo {author} {\bibfnamefont {T.}~\bibnamefont {Hayase}},
  \bibinfo {author} {\bibfnamefont {H.}~\bibnamefont {Kawamata}}, \bibinfo
  {author} {\bibfnamefont {K.}~\bibnamefont {Nakao}}, \bibinfo {author}
  {\bibfnamefont {A.}~\bibnamefont {Nakajima}}, \ and\ \bibinfo {author}
  {\bibfnamefont {K.}~\bibnamefont {Kaya}},\ }\href {\doibase 10.1063/1.471048}
  {\bibfield  {journal} {\bibinfo  {journal} {J. Chem. Phys.}\ }\textbf
  {\bibinfo {volume} {104}},\ \bibinfo {pages} {3413} (\bibinfo {year}
  {1996})}\BibitemShut {NoStop}%
\bibitem [{\citenamefont {Takano}\ \emph {et~al.}(2004)\citenamefont {Takano},
  \citenamefont {Yamamoto},\ and\ \citenamefont {Saito}}]{Takano2004}%
  \BibitemOpen
  \bibfield  {author} {\bibinfo {author} {\bibfnamefont {S.}~\bibnamefont
  {Takano}}, \bibinfo {author} {\bibfnamefont {S.}~\bibnamefont {Yamamoto}}, \
  and\ \bibinfo {author} {\bibfnamefont {S.}~\bibnamefont {Saito}},\ }\href
  {\doibase 10.1016/j.jms.2004.01.003} {\bibfield  {journal} {\bibinfo
  {journal} {J. Mol. Spectrosc.}\ }\textbf {\bibinfo {volume} {224}},\ \bibinfo
  {pages} {137} (\bibinfo {year} {2004})}\BibitemShut {NoStop}%
\bibitem [{\citenamefont {Drowart}\ \emph {et~al.}(1967)\citenamefont
  {Drowart}, \citenamefont {Pattoret},\ and\ \citenamefont
  {Smoes}}]{DrowartJ67}%
  \BibitemOpen
  \bibfield  {author} {\bibinfo {author} {\bibfnamefont {J.}~\bibnamefont
  {Drowart}}, \bibinfo {author} {\bibfnamefont {A.}~\bibnamefont {Pattoret}}, \
  and\ \bibinfo {author} {\bibfnamefont {S.}~\bibnamefont {Smoes}},\
  }\href@noop {} {\bibfield  {journal} {\bibinfo  {journal} {Proc. Brit. Ceram.
  Soc.}\ }\textbf {\bibinfo {volume} {8}},\ \bibinfo {pages} {67} (\bibinfo
  {year} {1967})}\BibitemShut {NoStop}%
\bibitem [{\citenamefont {Marquart}\ and\ \citenamefont
  {Berkowitz}(1963)}]{Marquart1963}%
  \BibitemOpen
  \bibfield  {author} {\bibinfo {author} {\bibfnamefont {J.~R.}\ \bibnamefont
  {Marquart}}\ and\ \bibinfo {author} {\bibfnamefont {J.}~\bibnamefont
  {Berkowitz}},\ }\href {\doibase 10.1063/1.1734242} {\bibfield  {journal}
  {\bibinfo  {journal} {J. Chem. Phys.}\ }\textbf {\bibinfo {volume} {39}},\
  \bibinfo {pages} {283} (\bibinfo {year} {1963})}\BibitemShut {NoStop}%
\bibitem [{\citenamefont {DeVore}\ and\ \citenamefont
  {Franzen}(1975)}]{DeVore1975}%
  \BibitemOpen
  \bibfield  {author} {\bibinfo {author} {\bibfnamefont {T.}~\bibnamefont
  {DeVore}}\ and\ \bibinfo {author} {\bibfnamefont {H.}~\bibnamefont
  {Franzen}},\ }\href@noop {} {\bibfield  {journal} {\bibinfo  {journal} {High
  Temp. Sci.}\ }\textbf {\bibinfo {volume} {7}} (\bibinfo {year}
  {1975})}\BibitemShut {NoStop}%
\bibitem [{\citenamefont {Zhai}\ \emph {et~al.}(2003)\citenamefont {Zhai},
  \citenamefont {Kiran},\ and\ \citenamefont {Wang}}]{Zhai2003}%
  \BibitemOpen
  \bibfield  {author} {\bibinfo {author} {\bibfnamefont {H.-J.}\ \bibnamefont
  {Zhai}}, \bibinfo {author} {\bibfnamefont {B.}~\bibnamefont {Kiran}}, \ and\
  \bibinfo {author} {\bibfnamefont {L.-S.}\ \bibnamefont {Wang}},\ }\href
  {\doibase 10.1002/chin.200326010} {\bibfield  {journal} {\bibinfo  {journal}
  {ChemInform}\ }\textbf {\bibinfo {volume} {34}},\ \bibinfo {pages} {2821}
  (\bibinfo {year} {2003})}\BibitemShut {NoStop}%
\bibitem [{\citenamefont {Wang}\ \emph {et~al.}(2011)\citenamefont {Wang},
  \citenamefont {Huang}, \citenamefont {Zhen}, \citenamefont {Zhang},\ and\
  \citenamefont {Chen}}]{Wang2011}%
  \BibitemOpen
  \bibfield  {author} {\bibinfo {author} {\bibfnamefont {L.}~\bibnamefont
  {Wang}}, \bibinfo {author} {\bibfnamefont {D.-l.}\ \bibnamefont {Huang}},
  \bibinfo {author} {\bibfnamefont {J.-f.}\ \bibnamefont {Zhen}}, \bibinfo
  {author} {\bibfnamefont {Q.}~\bibnamefont {Zhang}}, \ and\ \bibinfo {author}
  {\bibfnamefont {Y.}~\bibnamefont {Chen}},\ }\href {\doibase
  10.1088/1674-0068/24/01/1-3} {\bibfield  {journal} {\bibinfo  {journal}
  {Chinese J. Chem. Phys.}\ }\textbf {\bibinfo {volume} {24}},\ \bibinfo
  {pages} {1} (\bibinfo {year} {2011})}\BibitemShut {NoStop}%
\bibitem [{\citenamefont {H{\"{u}}bner}\ \emph {et~al.}(1998)\citenamefont
  {H{\"{u}}bner}, \citenamefont {Termath}, \citenamefont {Berning},\ and\
  \citenamefont {Sauer}}]{Hubner1998}%
  \BibitemOpen
  \bibfield  {author} {\bibinfo {author} {\bibfnamefont {O.}~\bibnamefont
  {H{\"{u}}bner}}, \bibinfo {author} {\bibfnamefont {V.}~\bibnamefont
  {Termath}}, \bibinfo {author} {\bibfnamefont {A.}~\bibnamefont {Berning}}, \
  and\ \bibinfo {author} {\bibfnamefont {J.}~\bibnamefont {Sauer}},\ }\href
  {\doibase 10.1016/S0009-2614(98)00792-1} {\bibfield  {journal} {\bibinfo
  {journal} {Chem. Phys. Lett.}\ }\textbf {\bibinfo {volume} {294}},\ \bibinfo
  {pages} {37} (\bibinfo {year} {1998})}\BibitemShut {NoStop}%
\bibitem [{\citenamefont {Bridgeman}\ and\ \citenamefont
  {Rothery}(2000)}]{Bridgeman2000}%
  \BibitemOpen
  \bibfield  {author} {\bibinfo {author} {\bibfnamefont {A.~J.}\ \bibnamefont
  {Bridgeman}}\ and\ \bibinfo {author} {\bibfnamefont {J.}~\bibnamefont
  {Rothery}},\ }\href {\doibase 10.1039/a906523g} {\bibfield  {journal}
  {\bibinfo  {journal} {J. Chem. Soc. Dalt. Trans.}\ ,\ \bibinfo {pages} {211}}
  (\bibinfo {year} {2000})}\BibitemShut {NoStop}%
\bibitem [{\citenamefont {Schultz}\ \emph {et~al.}(2005)\citenamefont
  {Schultz}, \citenamefont {Zhao},\ and\ \citenamefont
  {Truhlar}}]{Schultz2005a}%
  \BibitemOpen
  \bibfield  {author} {\bibinfo {author} {\bibfnamefont {N.~E.}\ \bibnamefont
  {Schultz}}, \bibinfo {author} {\bibfnamefont {Y.}~\bibnamefont {Zhao}}, \
  and\ \bibinfo {author} {\bibfnamefont {D.~G.}\ \bibnamefont {Truhlar}},\
  }\href {\doibase 10.1021/jp0539223} {\bibfield  {journal} {\bibinfo
  {journal} {J. Phys. Chem. A}\ }\textbf {\bibinfo {volume} {109}},\ \bibinfo
  {pages} {11127} (\bibinfo {year} {2005})}\BibitemShut {NoStop}%
\bibitem [{\citenamefont {Wu}\ \emph {et~al.}(2007)\citenamefont {Wu},
  \citenamefont {Wang},\ and\ \citenamefont {Su}}]{Wu2007}%
  \BibitemOpen
  \bibfield  {author} {\bibinfo {author} {\bibfnamefont {Z.~J.}\ \bibnamefont
  {Wu}}, \bibinfo {author} {\bibfnamefont {M.~Y.}\ \bibnamefont {Wang}}, \ and\
  \bibinfo {author} {\bibfnamefont {Z.~M.}\ \bibnamefont {Su}},\ }\href
  {\doibase 10.1002/jcc.20603} {\bibfield  {journal} {\bibinfo  {journal} {J.
  Comput. Chem.}\ }\textbf {\bibinfo {volume} {28}},\ \bibinfo {pages} {703}
  (\bibinfo {year} {2007})}\BibitemShut {NoStop}%
\bibitem [{\citenamefont {Clima}\ and\ \citenamefont
  {Hendrickx}(2007)}]{Clima2007}%
  \BibitemOpen
  \bibfield  {author} {\bibinfo {author} {\bibfnamefont {S.}~\bibnamefont
  {Clima}}\ and\ \bibinfo {author} {\bibfnamefont {M.~F.~A.}\ \bibnamefont
  {Hendrickx}},\ }\href {\doibase 10.1016/j.cplett.2007.01.073} {\bibfield
  {journal} {\bibinfo  {journal} {Chem. Phys. Lett.}\ }\textbf {\bibinfo
  {volume} {436}},\ \bibinfo {pages} {341} (\bibinfo {year}
  {2007})}\BibitemShut {NoStop}%
\bibitem [{\citenamefont {Liang}\ \emph {et~al.}(2009)\citenamefont {Liang},
  \citenamefont {Wang},\ and\ \citenamefont {Andrews}}]{Liang2009}%
  \BibitemOpen
  \bibfield  {author} {\bibinfo {author} {\bibfnamefont {B.}~\bibnamefont
  {Liang}}, \bibinfo {author} {\bibfnamefont {X.}~\bibnamefont {Wang}}, \ and\
  \bibinfo {author} {\bibfnamefont {L.}~\bibnamefont {Andrews}},\ }\href
  {\doibase 10.1021/jp900994c} {\bibfield  {journal} {\bibinfo  {journal} {J.
  Phys. Chem. A}\ }\textbf {\bibinfo {volume} {113}},\ \bibinfo {pages} {5375}
  (\bibinfo {year} {2009})}\BibitemShut {NoStop}%
\bibitem [{\citenamefont {Li}\ \emph {et~al.}(2013)\citenamefont {Li},
  \citenamefont {Wang}, \citenamefont {Wang}, \citenamefont {Gao},
  \citenamefont {Geng}, \citenamefont {Li}, \citenamefont {Wang},\ and\
  \citenamefont {Jiao}}]{Li2013}%
  \BibitemOpen
  \bibfield  {author} {\bibinfo {author} {\bibfnamefont {Y.-N.}\ \bibnamefont
  {Li}}, \bibinfo {author} {\bibfnamefont {S.}~\bibnamefont {Wang}}, \bibinfo
  {author} {\bibfnamefont {T.}~\bibnamefont {Wang}}, \bibinfo {author}
  {\bibfnamefont {R.}~\bibnamefont {Gao}}, \bibinfo {author} {\bibfnamefont
  {C.-Y.}\ \bibnamefont {Geng}}, \bibinfo {author} {\bibfnamefont {Y.-W.}\
  \bibnamefont {Li}}, \bibinfo {author} {\bibfnamefont {J.}~\bibnamefont
  {Wang}}, \ and\ \bibinfo {author} {\bibfnamefont {H.}~\bibnamefont {Jiao}},\
  }\href {\doibase 10.1002/cphc.201201043} {\bibfield  {journal} {\bibinfo
  {journal} {ChemPhysChem}\ }\textbf {\bibinfo {volume} {14}},\ \bibinfo
  {pages} {1182} (\bibinfo {year} {2013})}\BibitemShut {NoStop}%
\bibitem [{\citenamefont {Bauschlicher}\ and\ \citenamefont
  {Maitre}(1995)}]{Bau1995}%
  \BibitemOpen
  \bibfield  {author} {\bibinfo {author} {\bibfnamefont {C.~W.}\ \bibnamefont
  {Bauschlicher}}\ and\ \bibinfo {author} {\bibfnamefont {P.}~\bibnamefont
  {Maitre}},\ }\href {\doibase 10.1007/BF01113847} {\bibfield  {journal}
  {\bibinfo  {journal} {Theor. Chim. Acta}\ }\textbf {\bibinfo {volume} {90}},\
  \bibinfo {pages} {189} (\bibinfo {year} {1995})}\BibitemShut {NoStop}%
\bibitem [{\citenamefont {Schmidt}\ \emph {et~al.}(1993)\citenamefont
  {Schmidt}, \citenamefont {Baldridge}, \citenamefont {Boatz}, \citenamefont
  {Elbert}, \citenamefont {Gordon}, \citenamefont {Jensen}, \citenamefont
  {Koseki}, \citenamefont {Matsunaga}, \citenamefont {Nguyen}, \citenamefont
  {Su}, \citenamefont {Windus}, \citenamefont {Dupuis},\ and\ \citenamefont
  {Montgomery}}]{Gamess}%
  \BibitemOpen
  \bibfield  {author} {\bibinfo {author} {\bibfnamefont {M.~W.}\ \bibnamefont
  {Schmidt}}, \bibinfo {author} {\bibfnamefont {K.~K.}\ \bibnamefont
  {Baldridge}}, \bibinfo {author} {\bibfnamefont {J.~A.}\ \bibnamefont
  {Boatz}}, \bibinfo {author} {\bibfnamefont {S.~T.}\ \bibnamefont {Elbert}},
  \bibinfo {author} {\bibfnamefont {M.~S.}\ \bibnamefont {Gordon}}, \bibinfo
  {author} {\bibfnamefont {J.~H.}\ \bibnamefont {Jensen}}, \bibinfo {author}
  {\bibfnamefont {S.}~\bibnamefont {Koseki}}, \bibinfo {author} {\bibfnamefont
  {N.}~\bibnamefont {Matsunaga}}, \bibinfo {author} {\bibfnamefont {K.~A.}\
  \bibnamefont {Nguyen}}, \bibinfo {author} {\bibfnamefont {S.}~\bibnamefont
  {Su}}, \bibinfo {author} {\bibfnamefont {T.~L.}\ \bibnamefont {Windus}},
  \bibinfo {author} {\bibfnamefont {M.}~\bibnamefont {Dupuis}}, \ and\ \bibinfo
  {author} {\bibfnamefont {J.~A.}\ \bibnamefont {Montgomery}},\ }\href
  {\doibase 10.1002/jcc.540141112} {\bibfield  {journal} {\bibinfo  {journal}
  {Journal of Computational Chemistry}\ }\textbf {\bibinfo {volume} {14}},\
  \bibinfo {pages} {1347} (\bibinfo {year} {1993})}\BibitemShut {NoStop}%
\bibitem [{\citenamefont {Melton}\ \emph {et~al.}(2016)\citenamefont {Melton},
  \citenamefont {Zhu}, \citenamefont {Guo}, \citenamefont {Ambrosetti},
  \citenamefont {Pederiva},\ and\ \citenamefont {Mitas}}]{Melton2016}%
  \BibitemOpen
  \bibfield  {author} {\bibinfo {author} {\bibfnamefont {C.~A.}\ \bibnamefont
  {Melton}}, \bibinfo {author} {\bibfnamefont {M.}~\bibnamefont {Zhu}},
  \bibinfo {author} {\bibfnamefont {S.}~\bibnamefont {Guo}}, \bibinfo {author}
  {\bibfnamefont {A.}~\bibnamefont {Ambrosetti}}, \bibinfo {author}
  {\bibfnamefont {F.}~\bibnamefont {Pederiva}}, \ and\ \bibinfo {author}
  {\bibfnamefont {L.}~\bibnamefont {Mitas}},\ }\href {\doibase
  10.1103/PhysRevA.93.042502} {\bibfield  {journal} {\bibinfo  {journal} {Phys.
  Rev. A}\ }\textbf {\bibinfo {volume} {93}},\ \bibinfo {pages} {042502}
  (\bibinfo {year} {2016})}\BibitemShut {NoStop}%
\bibitem [{\citenamefont {Kramida}\ \emph {et~al.}(2015)\citenamefont
  {Kramida}, \citenamefont {{Yu.~Ralchenko}}, \citenamefont {Reader},\ and\
  \citenamefont {{and NIST ASD Team}}}]{NIST_ASD}%
  \BibitemOpen
  \bibfield  {author} {\bibinfo {author} {\bibfnamefont {A.}~\bibnamefont
  {Kramida}}, \bibinfo {author} {\bibnamefont {{Yu.~Ralchenko}}}, \bibinfo
  {author} {\bibfnamefont {J.}~\bibnamefont {Reader}}, \ and\ \bibinfo {author}
  {\bibnamefont {{and NIST ASD Team}}},\ }\href@noop {} {}\bibinfo
  {howpublished} {{NIST Atomic Spectra Database (ver. 5.3),
  {\tt{physics.nist.gov/asd}}}} (\bibinfo {year} {2015})\BibitemShut {NoStop}%
\bibitem [{\citenamefont {Nakano}(1993{\natexlab{a}})}]{Nakano19931}%
  \BibitemOpen
  \bibfield  {author} {\bibinfo {author} {\bibfnamefont {H.}~\bibnamefont
  {Nakano}},\ }\href {\doibase 10.1016/0009-2614(93)89016-B} {\bibfield
  {journal} {\bibinfo  {journal} {Chem. Phys. Lett.}\ }\textbf {\bibinfo
  {volume} {207}},\ \bibinfo {pages} {372} (\bibinfo {year}
  {1993}{\natexlab{a}})}\BibitemShut {NoStop}%
\bibitem [{\citenamefont {Nakano}(1993{\natexlab{b}})}]{Nakano1993}%
  \BibitemOpen
  \bibfield  {author} {\bibinfo {author} {\bibfnamefont {H.}~\bibnamefont
  {Nakano}},\ }\href {\doibase 10.1063/1.465674} {\bibfield  {journal}
  {\bibinfo  {journal} {J. Chem. Phys.}\ }\textbf {\bibinfo {volume} {99}},\
  \bibinfo {pages} {7983} (\bibinfo {year} {1993}{\natexlab{b}})}\BibitemShut
  {NoStop}%
\bibitem [{\citenamefont {Koga}\ \emph {et~al.}(1999)\citenamefont {Koga},
  \citenamefont {Tatewaki}, \citenamefont {Matsuyama},\ and\ \citenamefont
  {Satoh}}]{Koga1999}%
  \BibitemOpen
  \bibfield  {author} {\bibinfo {author} {\bibfnamefont {T.}~\bibnamefont
  {Koga}}, \bibinfo {author} {\bibfnamefont {H.}~\bibnamefont {Tatewaki}},
  \bibinfo {author} {\bibfnamefont {H.}~\bibnamefont {Matsuyama}}, \ and\
  \bibinfo {author} {\bibfnamefont {Y.}~\bibnamefont {Satoh}},\ }\href
  {\doibase 10.1007/s002140050479} {\bibfield  {journal} {\bibinfo  {journal}
  {Theor. Chem. Acc.}\ }\textbf {\bibinfo {volume} {102}},\ \bibinfo {pages}
  {105} (\bibinfo {year} {1999})}\BibitemShut {NoStop}%
\bibitem [{\citenamefont {Noro}\ \emph {et~al.}(2000)\citenamefont {Noro},
  \citenamefont {Sekiya}, \citenamefont {Koga},\ and\ \citenamefont
  {Matsuyama}}]{Noro2000}%
  \BibitemOpen
  \bibfield  {author} {\bibinfo {author} {\bibfnamefont {T.}~\bibnamefont
  {Noro}}, \bibinfo {author} {\bibfnamefont {M.}~\bibnamefont {Sekiya}},
  \bibinfo {author} {\bibfnamefont {T.}~\bibnamefont {Koga}}, \ and\ \bibinfo
  {author} {\bibfnamefont {H.}~\bibnamefont {Matsuyama}},\ }\href {\doibase
  10.1007/s002140000126} {\bibfield  {journal} {\bibinfo  {journal} {Theor.
  Chem. Acc.}\ }\textbf {\bibinfo {volume} {104}},\ \bibinfo {pages} {146}
  (\bibinfo {year} {2000})}\BibitemShut {NoStop}%
\end{thebibliography}%
\end{document}